\documentclass[secnumarabic,amssymb, nobibnotes, aps, prd]{revtex4}%
\usepackage{graphicx}
\usepackage{dcolumn}
\usepackage{bm}
\usepackage{amsmath}%
\usepackage{amsfonts}%
\usepackage{amssymb}
\begin{document}

\title{Search for muon enhancement at sea level from transient solar activity}

\author{C. R. A. Augusto}
\address{Instituto de F\'{\i}%
sica Universidade Federal Fluminense, 24210-130,
Niter\'{o}i, RJ, Brazil} 

\author{C. E. Navia}
\address{Instituto de F\'{\i}%
sica Universidade Federal Fluminense, 24210-130,
Niter\'{o}i, RJ, Brazil} 

\author{M. B. Robba}
\address{Instituto de F\'{\i}%
sica Universidade Federal Fluminense, 24210-130,
Niter\'{o}i, RJ, Brazil}

\date{\today}

\begin{abstract}
This paper presents first results of an ongoing study of a possible association
between muon enhancements at ground observed by the TUPI telescope and transient events 
such as the Sun's X-ray activity. The analysis of the observed 
phenomenon by using the GOES satellite archive data seems to indicate 
that on most cases the Ground Level Enhancements (GLEs) could potentially  
be associated with solar flares. 
We show that small scale solar flares, those with prompt X-ray emission 
classified as C class (power above $10^{-6}Watts\;m^{-2}$ at 1 AU) may give 
rise to GLEs, probably associated with solar protons and ions arriving
to the Earth as a coherent particle pulse.
The TUPI telescope's high performance with these energetic solar particles   
arises mainly from: (1) its high counting rate (up to $\sim 100$ KHz). This value in most 
cases is around 100 times higher than other detectors at ground and (2) due to its 
tracking system. The telescope is always looking  near the direction of the IMF lines.    
The GLE's delay in relation of the  X-ray prompt emission 
suggest that shock driven by corona mass ejection 
(CME) is an essential requirement for the particle acceleration efficiency.
\end{abstract}


\maketitle

\section{Introduction}

As a consequence of the interaction of galactic cosmic ray particles with air nuclei in the upper 
atmosphere a wide variety of secondary particles are produced. Most
of these particles are pions. They initiate an electromagnetic 
shower via $\pi^0$ production followed by the decay
$\pi^0 \to \gamma \gamma$ (called the soft component). The 
shower has a muon component mainly from $\pi^{\pm}$ production
followed by the decay $\pi^{\pm} \to \mu^{\pm}\nu_{\mu}$ 
(called the hard component).  The count rates of these particles by detectors 
at ground level is characterized mostly by the geomagnetic value of the site where the 
detector is placed.  The solar activity, including the transient solar events 
such as solar flares,
has a strong influence on the count rates of particles in the sub-GeV to GeV energy region 
and under certain conditions they give  origin to the so called ground level 
enhancements (GLEs). There is also a small fraction of
muons that have their origin in photo-nuclear reactions induced by primary gamma rays. 
The ability to distinguish these "photo-nuclear" muons from background muons
depends upon statistics, the strength and energy spectrum of the emitting
source and some characteristics of the experiment such as the angular
resolution. 

The TUPI telescope can detect muons at sea level
with energies greater than the $\sim 0.2-0.3$ GeV required to penetrate the two 
flagstone or walls surrounding the telescope (see Appendix A).
The TUPI muon telescope is sensitive to primary particles 
(including photons) with energies above the
pion production energy. In the case of charged particles, the minimal primary energy 
must be compatible with the (Niteroi-Brazil) geomagnetic cut-off (= 9.8 GV or 9.8 GeV for proton).
Due to its limited aperture (9.5 degrees of opening angle), the TUPI telescope
is on the boundary between telescopes with a very small field of view, like
the air Cherenkov telescopes, and the small air shower arrays, characterized
by a large field of view. 

Sunspot groups can trigger solar flares and coronal mass ejection 
(CME), and exhibit a 
complex magnetic field that harbors energy for powerful eruptions. Solar flares 
and CMEs occurs whenever there is a rapid large-scale change in the Sun's 
magnetic field. In general the solar active region that produced the eruption has
a complicated magnetic configuration.
So far, we have found reports of the detection at ground level of
GLEs associated to solar powerful flares of large scale, those with a X-ray   
prompt emission classified as X-class (above $10^{-4} W\;m^{-2}$)
\cite{bieber02,falcone99,swinson90}, as well as 
the enhancement of muons at ground level from powerful solar  
flares have been reported \cite{poirier02,munakata01}.
Observations in satellites and at ground of solar flares \cite{smart96} 
have led to the identification of two classes of acceleration events: impulsive (prompt)  
and gradual (post-eruptive or delayed). The impulsive events require selective acceleration   
such as the gyroresonant interaction with plasma waves. The energetic particles  
from these events arrive very quickly, around 15-25 minutes after a flare. In contrast, the gradual   
events have a strong association with coronal mass ejection (CME) and suggest that the particles   
in these events are accelerated by CME driven shocks. The energetic particles 
from these events  
are observed up to several hours after a flare. The effect on the interplanetary medium occurs
preferentially during this post-eruptive phase \cite{kocharov95}.

In order to see a possible photo-muon excess in the direction of
the galactic center, since June 2003 a set of observations have been conducted
using the TUPI telescope and several GLEs
have been found. Here, we present a summary of the results obtained from these
observations and a search for a possible correlation between these GLEs and
satellite transient events due to solar X-ray activity. We present several 
correlations between flares whose X-ray prompt emission are of small scale and muon 
enhancements at ground level.

This paper is organized as follows: In section 2 the TUPI muon telescope is presented,
the experimental setup and its sensitivity to small scale flares is briefly discussed.
In section 3 the methods of observation including
the raster search technique and pressure corrected relative intensity are presented.
The section 4 contains the justification for the criteria of association between GLEs and 
solar flares used in the present study.
The results of the search for the origin of the GLEs are shown in section 5, including 
some comments, and section 6 contains conclusions and remarks.

\section{The TUPI muon telescope}

\subsection{Experimental setup}

Figure 1 is a scheme of the TUPI muon telescope, it has an equatorial
assembly and a servo-mechanism which allows the axis of the telescope 
to be pointed so
as to accompany a given source \cite{augusto03}. The telescope computes mainly the muon
intensity in the atmosphere initiated by cosmic rays (mainly protons), giving
the coincidence counting rate of two-elements A and B (plastic scintillator
detectors with $0.5m\times0.5m\times35cm$) placed perpendicularly 
to the axis of the telescope. The main task of the first-level trigger is a 
coincidence between the A and B detectors. The second-level trigger is a veto for air 
showers coming  from other directions, far off the telescope's axis direction and 
contains other two detectors (C and D Plastic scintillator) off the telescope axis.
Each plastic scintillator is viewed by a 7 cm diameter
photomultiplier and the unit has a nominal energy threshold of $\sim10MeV$ for
charged particles. With these characteristics the telescope has a effective 
angular aperture or geometrical factor  of $65.5\;cm^2sr$,  projecting to the space a cone  
with an open angle of $9.5^0$ in relation to the telescope's axis.

The TUPI muon telescope is installed on the campus of the Universidade Federal
Fluminense, Niter\'oi, Rio de Janeiro-Brazil. The position is: latitude:
$22^{0}54^{\prime}33^{\prime\prime}$ S, longitude: $43^{0}08^{\prime
}39^{\prime\prime}$ W, at sea level.
At sea level and in the sub-GeV to GeV energy region, the muon's flux is   
$\sim 70$ times higher than the nucleons flux, and $\sim 800$ times higher than the   
electrons' flux \cite{gaisser04}. Consequently the telescope detect mainly
muons with energies greater than 
the $\sim 0.2-0.3$ GeV required to penetrate the two flagstones or walls surrounding 
the telescope . The concrete reduces the noise due to other non-muon 
particles, for example it is opaque to electrons (see Appendix A for details). 
Outside of the building where the TUPI telescope is deployed, there is an
automatic meteorological station and we have free access to the data. In addition,
we have also our own barometer. Together with the muon counting rate, the pressure 
is registered every 10 second. The station is very useful because its data is
used as reference and calibration and other parameter as the outside temperature,
atmospheric humidity, and wind speed, among others are available.  

The data acquisition is made on the basis of an
Advantech PCI-1711/73 card with a counting rate up to 100 kHz.
All the steps from signal discrimination to the coincidence and anti-coincidence 
are made via software, using the virtual instrument technique. The application 
programs were written using the Lab-View tools.

\subsection{The sensitivity of the TUPI muon telescope}

It is commonly accept that a solar flare must be powerful 
to be a candidate for association with a GLE. Initially we do not 
believe that small-scale flares such as of the C class  could cause 
muon enhancements. However, now we have other opinion and in this section, 
we will try to respond to the question,  
which are the main factors that increase the TUPI telescope's sensitivity
to small scale flares?  

(a)The GLEs as observed by the TUPI telescope at sea level are characterized 
by an impulsive peak with a fast rise time (see section 5). This is a signature indicating 
that they are constituted by a bundle of muons produced in the atmosphere by the arrival   
of a coherent particle pulse, probably from the sun (see section 4).   
This mean that the particles from the pulse front arrive at the earth   
almost simultaneously. In order to discriminate (to count) a small coherent particle pulse, 
a detector working with a rather fast response time, better than milliseconds, is necessary
in order to reduce the dead time, during which the detector may not respond to the incident
radiation. The TUPI telescope  has a counting rate  up to $\sim 100$ KHz and at 
this frequency it is possible to obtain a response time as short as $\Delta T \sim 10 \;\mu s$. 
Table I shows the characteristics of several "solar particles" detectors at ground level
\cite{izmiran},
including the TUPI telescope. From Table I it is possible to see that typical 
neutron monitors (NM), NM-64, NM-scintillator and IGY types, have operated with a 
counting rate of $\sim 1$ KHZ.

(b) Due to the TUPI telescope's tracking system, the pitch angle
defined as the angle between the sun-ward direction and the telescope axis direction,
is always the same (constant) during a raster scan.
This characteristic probably also helps to increases the sensitivity 
because the energetic solar charged particle propagation follows the 
interplanetary magnetic field lines (IMF), the rotation of the Sun gives the magnetic field 
a spiral form (garden hose effect)
the pitch angle of the IMF at 1 AU is $\sim 45^0$ (this value doesn't include the magnetic field 
of the Earth).
If the telescope axis is oriented near or close to the direction of these IMF lines, as
is schematized in Fig.2 (see also the section 4),
the solar particle sources will be magnetically well connected to the direction of the
telescope axis and the detection efficiency will be close to maximum. This favorable
situation increases the telescope's sensitivity.
 
\section{Method of observation} 

\subsection{Raster search technique}

The atmospheric muon flux originating from the decay of charged pions and kaons  
produced by galactic primary cosmic rays in the atmosphere constitute the main
source of background at sea level and in the energy region of sub-GeV to GeV.
However, at these energies the muon flux 
is influenced by the magnetic field of the Earth \cite{hayakawa69}.
Consequently the muon angular distributions  are
quite different for different sites on the globe.
There are two main geomagnetic effects on the muon flux observed at ground level:
a) The East-West effect, the muon flux is highest (lowest) for a direction coming from 
the West (East). b)The azimuth dependence on the positive-to-negative ratio of muons, 
a considerable amount of the negative excess is observed for muons coming from the east, 
because a positive muon coming from the East has a longer path length than a negative one.   
These geomagnetic effects distort the zenith angle distribution of sub-GeV to GeV muons
during a raster scan, because the measures (galactic center) begin around the South-East 
direction and they finish around the South-West to southerly 
directions.

The observation of the galactic center ($Declination=-29^{0}$,
$RightAscension=17h,42m$) began on June 26, 2003 and consists of 
on-source and off-source tracking runs or raster scans, where each
tracking run represents a sidereal half-day (12 hours). This is approximately
the time that the galactic center is above our horizon in every 24 hours. 
Basically each tracking
run is a time series which is generated registering the coincident counting
rate every 20 seconds and initially in order to avoid the background contamination
a very high (pulse-height) discrimination level has been used. 
However, starting from 2003 December, after a up-grade, the counter rate 
every 10 seconds and a low pulse-height discrimination have been used to observe 
a possible muon excess in the direction of the galactic center.  

Until February 28, 2005  we have  completed 2160 hour of observation which constituted
180 raster scans (12 hour each). In this paper we are presenting the first six
GLEs, All of them have the 
potential to be considered as solar flare associated, because there is
some correlation with solar flares, whose prompt X-ray emission have been
reported by the GOES spacecraft (NOAA Space Environment Center Website)\cite{noaa}.
In all cases except one, the GLEs were detected with a statistical
significance of more than $3\sigma$ and they are described in section 5.

\subsection{Pressure correction and relative intensity}

In order to taken into account the barometric pressure effect in the muon intensity,
the muon counting rate is transformed to the "relative intensity" by using the Natural 
Logarithmic Representation (NLR). The algorithm comes from the exponential dependence 
of the counting rate on the pressure
\begin{equation}
N=N_0\times \exp[-\beta(P-P_0)],
\end{equation}
where $\beta$ is the barometric coefficient. It has a value of $\sim 1\%$ per mbar
(in neutron monitors experiments). However, for the muon component, $\beta$ has a value 
below $0.7\%/mbar$, varying from station to station.
$P_0$ is the reference's pressure (average pressure). 
Fig.3 shows the hourly average distribution of the atmospheric pressure, obtained in 2003
at Niteroi city with $P_0(=1014.73\pm4.75)\;mbar$. $N_0$ is the average counting rate.
In the NRL representation the relative intensity is defined as
\begin{equation}
W=10^2 \times \ln (N/N_0), \;\;in\;units\;of\;\%.
\end{equation}
In this way, the pressure correction on the relative intensity is given by
\begin{equation}
W_p=W-\beta \times (P-P_0).
\end{equation}
 Following the last relation, it is possible to see
that the relative muon intensity is in $anti-correlation$ with the pressure, or in other words,
if the atmospheric pressure $P$ is larger than the average value $P_0$ the relative muon 
intensity decreases and vise versa. In this paper, every raster scan (raw data) is divided in bins 
of 15 minutes and  then we apply the pressure correction algorithm. 
From Fig.3 it is possible to see that the bandwidth of the pressure variation is around
$\Delta P \sim \pm 4.8$ mbar (at 95\% confidence level) and correspond to a variation in the muon relative 
intensity of up to $\sim \pm 3.4\%$

\section{GLE associated flare}

The occurrence of GLEs (muons' excess) as observed by the TUPI 
telescope can be interpreted  as evidence of the arrival in the upper
layers of the atmosphere of a bundle of protons and/or ions with 
energies exceeding the pion production and above the local geomagnetic cut-off 
(9.8 GV).
However, the central question is: are they associated with small-scale flares?
In the affirmative case, what are the acceleration mechanisms to GeV energies?
In order to give a convincing answer to these questions an analysis
on the basis of spacecraft data was made.

The statistical studies on the association between the SOHO/LASCO CMEs 
and the GOES X-ray flares provides us with useful tools to the analysis
on the nature of the GLEs apparently associated with small X-ray flares.
In the case of gradual events, whose origin results from the acceleration of 
ambient coronal ions (i.e. protons, alpha etc)
by shocks driven by CMEs, a plausible explanation (among others) for the 
low rate of GLEs associated to small scale solar flares 
is the so called  "filtering effect"
\cite{aoki04}. This effect has been obtained from
recent solar observations on the basis of the SOHO/LASCO CME CATALOG and the
Solar Geophysical Data (SGD) for the same period. The results 
have reveled that mass ejection are ubiquitous in solar flares. However, if the
flare is less energetic, the hot plasma is surrounded by a strong magnetic 
field and cannot escape from the corona. In this case there is not CME ejected to 
the interplanetary space, there is only the X-ray emission. 
This effect can explain the shape of the distribution of the X-ray peak fluxes of 
CME-associated flares and that it is of the type Log-normal distribution.
On the other hands, the distribution of the number of flares per unit 
energy and per unit time follows a power law $N(E) \sim E^{-\delta}$, with
$\delta \sim 1.5$ and this index practically does not vary with the solar cycle
\cite{crosby93}. 
Consequently the number of small flares is high, while the CMEs associated with 
these small flares is low. They are in the left tail of the Log-normal distribution 
of the X-ray peak flux of CME-associated flares. The mean value 
(peak of the Log-normal distribution) is around  $6.81\times 10^{-6}Wm^{-2}$. 

The above results are compatible with other statistical studies of solar flares and CMEs.
Particularly with the waiting-time distribution, the distribution of times
$\Delta T$ between events \cite{wheatland03}. Both the solar flare and CME waiting-time distributions
are closely related by a Poisson distribution. The comparison between the waiting-time
distribution for LASCO CMEs and for GOES X-ray flares of greater than C1 class for
the years 1996-2001 has shown an excess of flares in relation to the CMES in the 
waiting-time region up to  $\Delta T \sim 2$ hours. This is exactly the region 
of small scale flares.

The nature of solar flares 
associated with GLEs via coronal mass ejection is poorly understood. Solar flares 
and CMEs often occur together, but not necessarily because the flare triggers the 
CME or vice versa. Especially during solar maximum, there are CMEs 
without an associated flare. However, the effect can be due to high solar activity,
because the X-ray background is hight and the flares with a small scale X-ray emission
can be masked, while small scale flares without an associated CMEs are 
more frequents due to the "filtering effect".

The conditions under which a CME is associated with a flare are found by addressing the 
relationships between their positions and timing the flare start and the appearance 
of  the CME. The association of events between SOHO/LASCO CMEs and GOES-8 X-ray flares
(band-width $1.0-8.0\;A$) \cite{shanmugaraju03}
have shown that the CMEs not associated with metric radio burst 
(CMEs no metric type II) have a delay in relation to the flare start of 
$30\pm 43.3\; min$). The X-ray peak flux of these
flares have a mean value of $6.5\times 10^{-6} Wm^{-2}$, consequently most 
of them are similar to C class. The CMEs such as metric type II are associated mostly 
with flares of M class.

The association between a X-ray flare and a GLE requires taking into account the 
delay between the flare start and appearance of CME plus the time of flight between the 
Sun and the Earth of highly energetic particles.
The time of flight is estimated from results of simulations under the assumption that
the propagation of energetic particles released (for instance) by a CME driven shocks
are injected into the interplanetary medium in coherent pulses of energetic particles
and a realistic Archimedean spiral field lines around the Sun \cite{ruffolo95}.
Here, we summarized simulation results
for a typical arc length along the magnetic field (garden hose direction), 
$<z>(=1.3\;AU)$, as a function of the distance traveled, $S$, which can be expressed as
\begin{equation}
<z>=\alpha(\lambda) S.
\end{equation}
The constant of proportionality, $\alpha(\lambda)$, depend on the scattering mean free path, 
$\lambda$, and satisfies the constrain condition
$\alpha(\lambda)=1$ for $\lambda=1.0 AU$, sometimes called "scatter-free" condition and 
implies that particles freely stream along the field at their maximum speed like a coherent
particle pulse.

There are two extreme cases for high energy solar particle propagation to $1$ AU,
depending on how the particles are injected into the interplanetary medium:
When the scattering mean free path is very small compared to the scale length of the 
IMF(i.e.$\lambda \leq 0.3AU$). The particles propagation follows helical trajectories 
around the field of the IMF. Fluctuations of small scale of the IMF act as scattering 
centers of the particles and the propagation is dominantly diffusive. In contrast, 
when the scattering  mean free path is compatible with the distance from particles source 
(i.e. $\lambda \sim 1AU)$ the focusing effect of the Interplanetary Magnetic Field lines
became dominant and the propagation of the energetic particles is like coherent pulses,
following trajectories around the field line of the IMF.

GLEs constituted by muon bundles, detected at sea level by a directional telescope,
are a signature of primary particles arriving at the top of the atmosphere with a strong
anisotropic pitch angle distribution "coherent pulses", as well as, 
the very short rise time in the profile time of the GLEs suggest a non-diffusive 
coherent particle pulse transport. Under the non-diffusive transport assumption and 
for a typical average value of $\lambda=0.9\;AU$, the time of flight
of coherent pulses of energetic particles with a mean
rigidity of $10 GV$ is  estimated as $35\pm 51$ minutes.

Consequently, we look for the more energetic (larger peak flux) and nearest X-ray 
flare from the GLEs with a delay of $1.08\pm 1.57$ hours. The criteria comes from the 
"filtering effect", because the probability of occur of a CME associated flare
increase with the energy. On the other hand, the $1.08\pm 1.25$ hours window take into account 
the delay between flare start and appearance of CME plus the time of flight between 
the Sun and the Earth of highly energetic particles.

On the other hand, even in GLEs associated with large 
solar flares, the acceleration mechanism producing particles of up to several tens of 
GeV is not well understood. The situation becomes still more critic in the case 
of GLEs associated with solar flares of small scale. Here, we argue the possibility
of a "scale-free" energy distribution of particles accelerated by shock
linked with CMEs. This possibility comes from the shapes of the energy spectrum 
of the particles being close to a power-law distribution, 
because power law distributions are characterized as scale free distributions.
The spectral index of these power-law distributions depend on time t in one and the 
same event and of event to event due to stochastic processes of acceleration and 
modulation \cite{dvoraikov95}. The energy spectrum of energetic particles (protons and ions) during the 
early stage of a GLE is less steep than the energy spectrum at declining stage 
and probably the tail of the distributions extend to energies above 
20 GeV/nucleon. This scale-free hypothesis need further confirmation.  

\section{RESULTS}

We begin with a list (Table II) of all associated GLEs and solar flares
established in the present study, according to the criteria established in the last 
section. In the Table II, a zero degree pitch angle represents the sun-ward 
direction and the coordinates Right Ascension and Declination and pitch angle are
with reference to the telescope axis. Therefore the angular aperture of the
telescope is a cone of opening angle 9.5 degrees.
The plots and data files of the GLEs reported here are accessible to the public
through our web-site \cite{navia} (see results).
In second place we show the correlation between the X-ray solar emission
and the TUPI counting rate when there are no
solar flares (X-Ray prompt emission). In this case,
the solar X-ray emission (background) as reported by the GOES spacecraft
on $2004/09/28-29$ is as shown in Fig.4 (upper panel), the X-ray emission 
(band-width $1.0-8.0$ A) is 
lower than $10^{-6}Watts\;m^{-2}$ (the horizontal dash-dotted line shows this limit). 
In Fig.4 (central and lower panels), the time profiles of the telescope 
output (raw data) and the relative intensity (pressure corrected) are also presented.

In spite of the telescope's relatively low angular resolution ($9.5^0)$ and
the small statistics during a raster scan (12 hours),
it is possible to see the characteristics indicated above,   
such as a muon flux excess from the West direction. The West-East asymmetry is
defined as the (normalized) difference between the fluxes $(W-E)/(W+E)$ and is free
of calibrations and experimental bias. 
A quantitative result of the West-East effect during a raster
scan can be obtained from two symmetrical points, those with  the same zenith angle,
with one pointing to the West and the other to the East (see the two arrows
in Fig.4)). In the present case we obtain a West-East asymmetry around 16\%. 
This value is in agreement with similar observations \cite{tokiwa03}.

In figures 5, 6, 7, 8, 9 and 10 the time profiles of the GOES12 X-ray 
in the band-width $1.0-8.0$ A and $0.5-4.0$ A (upper panel), the TUPI raw data 
(central panel) and the TUPI relative intensity (in the NLR representation) pressure 
corrected (lower panel) are shown for six raster scans respectively. 
When a GLE  has been correlated with a solar flare, the X-ray peak 
(in the $1.0-8.0$ A band-width) of the flare 
is marked  by a flare index. For example, C1.3 means $1.3\times 10^{-6}Wm^{-2}$.    

We would like to make some comments on the experimental data:  
(a) The GLE found on $2003/10/23$ is classified 
as marginal due to the low statistical significance level ($\sim 2.2 \sigma$) 
in the raw data. However, a clear muon excess in the relative intensity 
(after pressure corrections) can be seen
It is practically in coincidence with a powerful X-ray ($X1.0$ class) prompt 
emission. The beginning of the GLE coincides approximately with the decline of the Sun.
Due to the small difference between the X-ray prompt emission and the GLE's peak,
the event can be classified as impulsive. The impulsive emission 
of powerful flares (such as those of X class) in most of the cases is observed 
at ground level by the network neutron monitors.
In all the other cases the delay of the GLEs in relation to the flares suggest,
emission via CMEs. CMEs produce fast shocks which accelerate charged particles
(electrons, protons and ions) in the magnetically open corona, over a region at
least comparable to the size of the CME itself. 

(b)In the occurrence of the GLE on 2003/12/02 especially the second peak $\sim 23h$ UT, the Sun
was below the horizon. Even so under this condition there is a magnetically well
connected field line. In the opposite case, and for low altitude of the Sun (early morning), 
the IMF are not well connected with the point of observation on the Earth. Fig.2 summarized 
the situation.

(c)Due to virtual instrument technique used in the data acquisition system,  
it is possible to examine the time profile of a GLE for other off-line (higher) pulse-height  
amplitude discrimination, chosen via software \cite{navia04}. Using  this "noise filter"  
we have verified that the six GLEs here presented  are not mere background fluctuations.

(d)If the Earth's magnetic field is ignored, the best value in the pitch angle to observe solar 
energetic particles is $45^0$ (see Fig.2). However, due to the geomagnetic effects, the pitch angle 
should be a little bigger (around 15\%-20\%). Due to the complexity of the phenomenon a full Monte Carlo  
is necessary in order to taking into account the effects of the joining of the IMF with the Earth's  
magnetic field on the propagation of the solar particles and their secondary particles produced   
in the atmosphere. The six GLEs analyzed here were obtained during an search of a possible excess   
of "phot-muons" in the direction of the galactic center, with a pitch angle bigger than that $45^0$.

%%%%%%%%%%%%%%%%%%%%%%%%%%%%%%%%%%%%%%%%%%

\section{Conclusions}

We have described and analysed  ground level enhancements
observed during a search for enhancements from the galactic center
with "photo-muons" at sea level, detected using the TUPI telescope. 
The paper present the first results of an ongoing study of the associations between
these GLEs and solar flares.
The main conclusions are summarized as follows:

(a)The TUPI muon telescope is sensitive to primary particles 
(including photons) with energies above the pion production threshold.
The TUPI telescope can detect muons at sea level with energies greater than 
the $0.2-0.3$ GeV required to penetrate the two flagstones or walls surrounding 
the telescope. The concrete reduces the noise due to other non-muonic particles.
For example, it is opaque to electrons.

b)The muon flux is subject
to several sources of modulation, such as the atmospheric pressure 
variation, and geomagnetic effect, among others. However, 
the temporal scale of these modulation phenomena are much larger than 
the GLEs duration. In addition, no anomalous changes in the atmospheric pressure
and temperature were observed during the raster scan where the GLEs were found. 
 
c) The impulsive emission of powerful flares (such as those of X class) 
is almost always accompanied by observations at ground level by the 
network neutron monitors, the delay between the X-ray prompt emission and 
the beginning of the GLE being
around ($15-30$ minutes). The GLE $2003/10/23$ has this characteristic. However,
the other  ground level enhancements have a delay of $\sim 1.5$ hours in 
relation to the  X-ray prompt emission and suggest an association with 
gradual or post-eruptive acceleration processes. Consequently, the shock driven by  
CME is an essential requirement for the particle acceleration efficiency.

d) The TUPI telescope has  shown the ability to detect variations in the muon fluxes.   
For  instance, it is possible to see the West-East effect embedded in the time series
(raw data). The efficiency of the TUPI telescope for detecting the enhancement of 
muons at sea level from a coherent cluster of solar energetic particles, associated 
to solar flares of small scale  is a consequence of several factors such as: (1) Its 
high counting rate, $\sim 100$ KHz. This value is around 100 times higher than the other 
telescopes at ground (see Table I). The requirement of a high counting rate is essential 
to discrimination of particles inside a small coherent particle pulse. (2)Due to its 
raster scan  system, the telescope axis can be oriented close to the directions of the 
IMF lines, favoring the detection of the solar particles.
(3)The shape of the lateral distribution function of muons at sea level, which is close 
to a flat distribution. For primary energies below 100 GeV \cite{poirier02a},
the fraction of muons that hit the telescope when the core of 
the shower is at a distance $r$ from the telescope center is the same for $r$ up to
$\sim 1.5$ km. 

In short, we have shown strong experimental evidences of the association between small scale 
flares and GLEs, but important questions as acceleration mechanism  of the solar particles 
up to GeV energies  remain open and requires continued observations and further investigation.
The ongoing experiment will deliver much better statistics in the next years. A careful 
study of the origin of the GLEs will be continued.

\begin{acknowledgments}
This work was partially supported by FAPERJ (Research foundation of the State
of Rio de Janeiro) in Brazil. The authors wish to express their thanks to Dr.
A. Ohsawa from Tokyo University for help in the first stage of the
experiment, to Dr. M. Olsen for reading the manuscript and to Dr. J.L.Fernandes de 
Oliveira from UFF University for the free access to the meteorological station data.
We are also grateful to the various catalogs available  on the internet and to 
their open data police, especially to the NOAA's Space Environment Center (SEC).
\end{acknowledgments}

\appendix
\section{Continuous charged lepton energy loss}
The average muon energy loss rate \cite{bichsel04} in its propagation through the matter is given by
\begin{equation}
-<\frac{dE}{dX}>=\alpha (E) +\beta (E) \times E,
\end{equation}
where X is the material thickness (in units of $g/cm^2$). The first term of the last 
equation (with $\alpha (E)\sim 2.0 MeV/g/cm^2$) take into account the ionization energy 
loss and the second term (with $\beta (E) \sim 6.0 \times 10^{-6}\;g/cm^2 $) represent 
the energy loss  due to radiative processes, and it is the dominant process only in 
the energy region above the critical energy (several hundred GeV). While for sub-GeV
to GeV energies the ionization became the dominant process to the energy loss
of muons. Under this assumption the last equation can be written as
\begin{equation}
-\Delta E \approx  2.0\times \Delta X \; MeV/g/cm^2.
\end{equation}
The TUPI telescope is inside a building, under two flagstones of concrete as is shown
in Fig.1. The flagstones' thickness have a dependence on the zenithal angle, with a 
average value of $\Delta X \sim 150\;g/cm^2$. Consequently,
the flagstones increases the detection muon energy threshold, because the telescope
will can detect muons with energies greater than the 0.2-0.3 GeV required to 
penetrate the two flagstones.

In the case of electrons the energy loss by bremsstrahlung become dominant above a 
few tens of MeV and is nearly independent on energy. For electrons of
100 to 1000 MeV, the energy loss can be approximately expressed as
\begin{equation}
-\frac{\Delta E}{\Delta X} \sim 0.15 (g/cm^2)^{-1} \times E(MeV).
\end{equation}
With the continuous and fast decreasing of the electrons'
energy via bremsstrahlung, and for energies below the critical value ($\sim 20 MeV$)
the ionization process become more important and they are "absorbed" by the flagstones
whose thickness is $\sim 150 g/cm^2$.
This mean that the two flagstones are opaque to the electrons.

\clearpage

\begin{widetext}
\begin{table}
\caption{Characteristic  of some "solar particle" detetores.}
\begin{ruledtabular}
\begin{tabular}{ccccc}
 & \multicolumn{3}{c}{Muon telescopes at ground} \\
\hline
 Experiment & Location (deg) & Altitude(m) & Counting Rate(KHz) & Rigidity cutoff(GV)\\
\hline
Nagoya(multi-directional)  & 35.2N,136.9E & 77 & 0.16 & 11.5  \\
TUPI (tracking system)    & 22.9S,43.3W & 3 & $\sim 100$   & 9.8  \\
GRAND (muon extensive air shower)   & 41.7N,86.2W &220 & 1.75     & 2.2 \\ 
\hline
\hline
& \multicolumn{3}{c}{Neutron Monitors} \\
\hline
Experiment & Location (deg) & Altitude(m) & Counting Rate(KHz) & Rigidity cutoff(GV)\\
\hline
Yangbaging (NM-64)  & 30.1N,90.5E & 4300 & 2.9 & 14.1  \\
Chacaltaya (Scintillator)  & 16.3S,221.8 & 5220 & $\sim 1.2$ & 13.1 \\
Climax (IGY) & 39.4N,253.8E&3400 & 4.3   & 2.99 \\
Haleakala (IGY)& 20.7N,203.7E&3030& 1.5  & 12.91 \\
\end{tabular}
\end{ruledtabular}
\end{table}

\end{widetext}

\begin{widetext}
\begin{table}
\caption{Chronology and main characteristics of the ground level enhancements.}
\begin{ruledtabular}
\begin{tabular}{cccccccc}
& \multicolumn{4}{c}{TUPI Telescope} &\vline&
\multicolumn{2}{c}{Pitch angle and delay in relation to the X-ray emission}\\
\hline
 Data & Start UT(h) & Significance level & RA(deg) & Dec(deg) & \vline & Pitch angle (deg) & delay (h) \\
\hline
2003/08/21  & 18.79 & $4.1 \sigma$ & 258.7 & -29  &\vline & 111 & 1.38 \\
2003/10/23  & 19.96 & $2.0 \sigma$ & 318.7 & -29  &\vline & 102 & 0.23  \\
2003/12/02  & 23:02 & $6.6 \sigma$ & 273.7 & -29  &\vline & Sun below horizon  & 1.50 \\
2004/09/10  & 18.32 & $14.4 \sigma$ & 258.7 & -29 &\vline & 91 & 1.63\\
2004/09/14  & 15.50 & $4.7 \sigma$ & 258.7 & -29  &\vline & 88 & 1.60 \\
2004/10/21  & 18.02 & $4.4 \sigma$ & 258.7 & -29  &\vline & 52 & 2.5 \\
\end{tabular}
\end{ruledtabular}
\end{table}

\end{widetext}
%%%%%%%%%%%%%%%%%%%%%%%%%%%%%%%%%%%%%%%%%%%%%%%%%%%%%%%%%%%%%%%%%%%%%%%%%%%%%%%%%%%%%%%% 

\begin{figure}[th]
\includegraphics[clip,width=1.2
\textwidth,height=1.0\textheight,angle=0.] {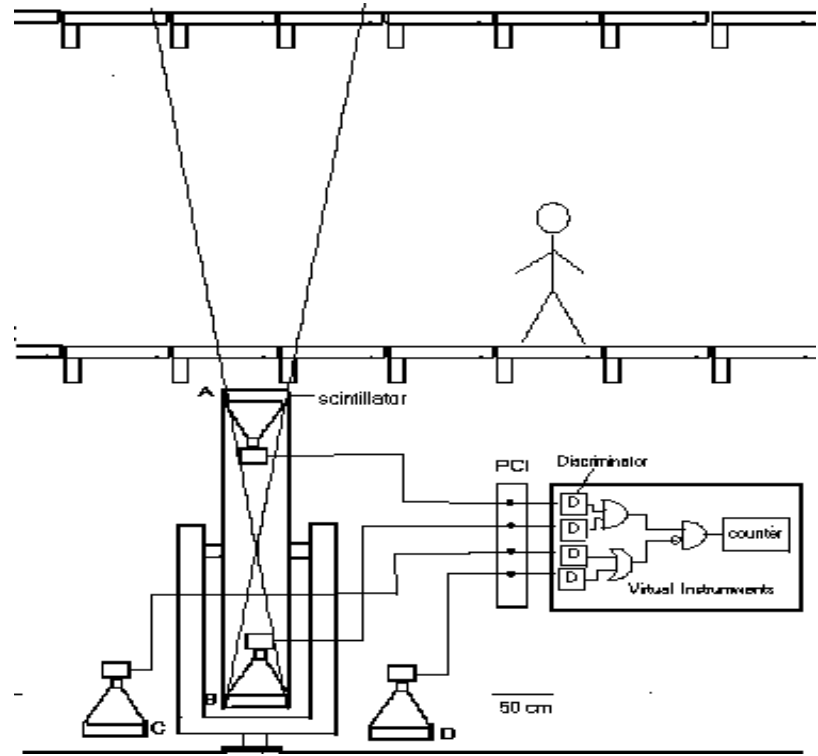}
\caption{Scheme of the TUPI muon telescope and the data acquisition system.}%
\end{figure}

\begin{figure}[th]
\includegraphics[clip,width=0.6
\textwidth,height=0.6\textheight,angle=0.] {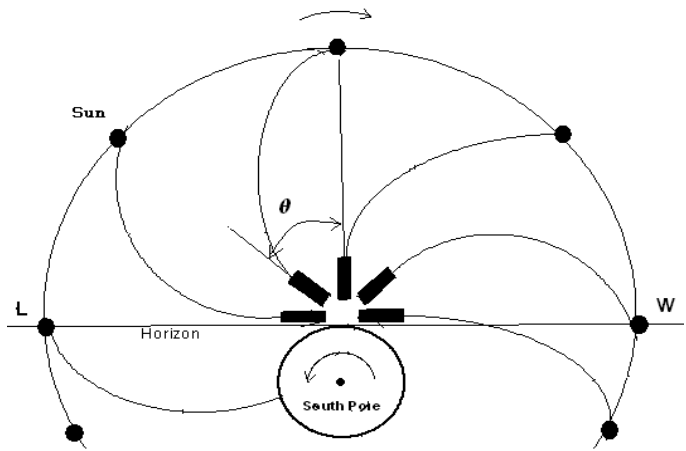}
\caption{Schematic representation in the South hemisphere of the IMF "garden hose" lines connecting 
the Sun and the Earth during a raster scan. Due to its tracking system the TUPI telescope axis is always 
close to the direction of the IMF. The scheme is for the best condition, $45^0$ of pitch angle
(see section 5).}%
\end{figure}

\begin{figure}[th]
\includegraphics[clip,width=0.6
\textwidth,height=0.6\textheight,angle=0.] {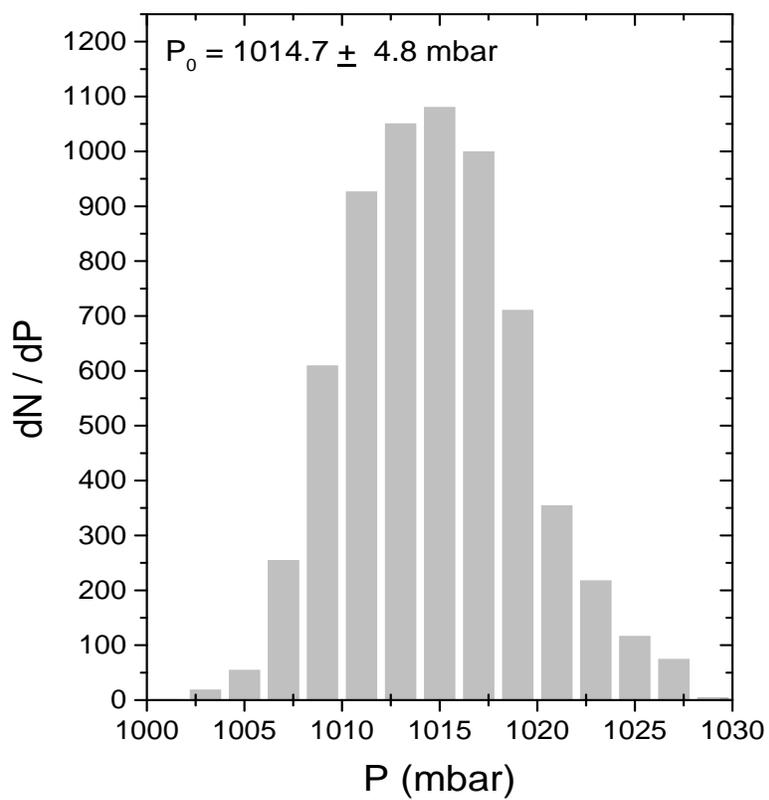}
\caption{Hourly average distribution of the atmospheric pressure, obtained in 2003 at 
Niteroi city.}%
\end{figure}

\begin{figure}[th]
\includegraphics[clip,width=0.6
\textwidth,height=0.6\textheight,angle=0.] {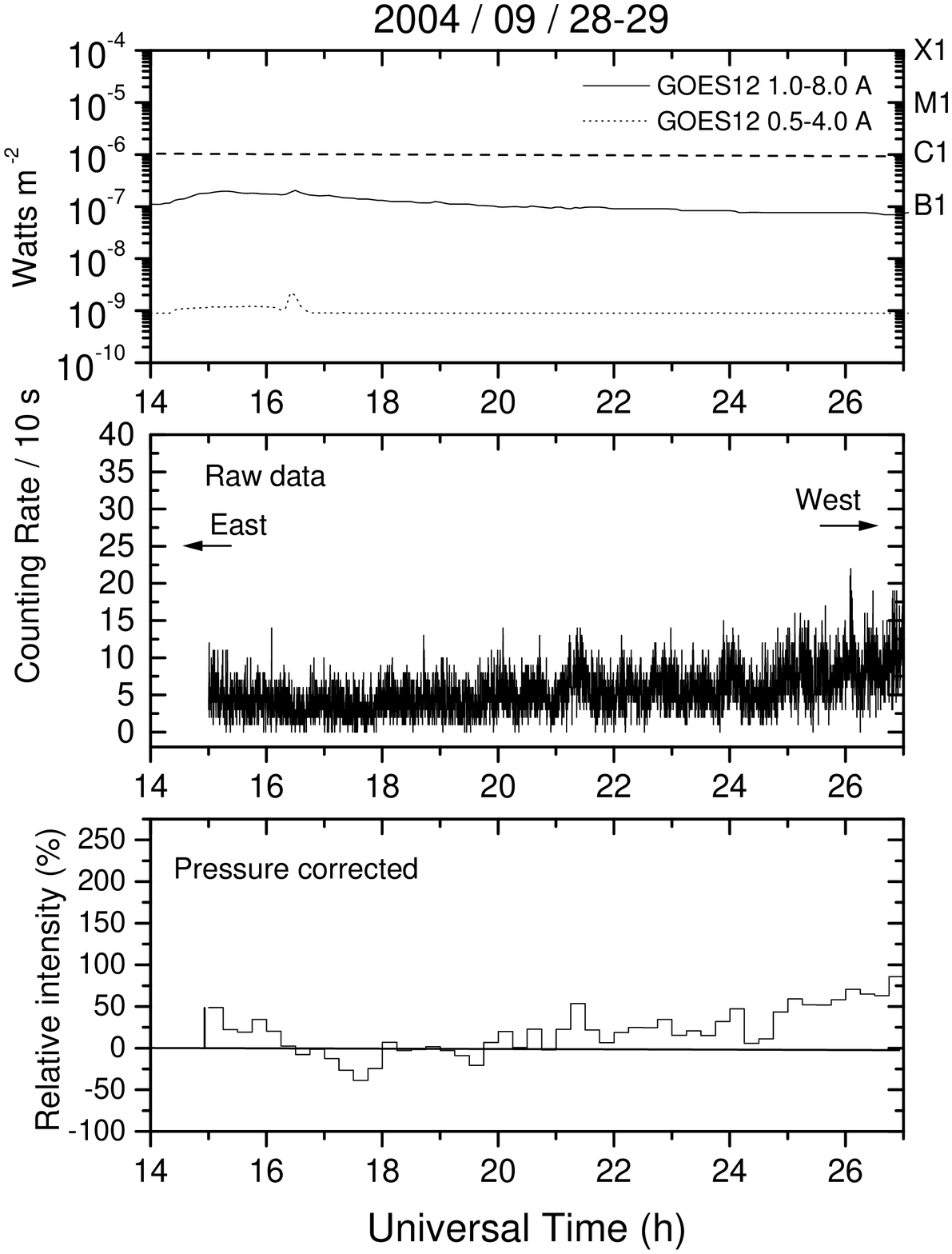}
\caption{Time profiles of the solar X-ray emission (background),
the TUPI raw data (counting rate), and the muon relative intensity (in the natural logarithmic representation) after pressure correction for the 2004/09/28-29  raster scan.}%
\end{figure}

\begin{figure}[th]
\includegraphics[clip,width=0.6
\textwidth,height=0.6\textheight,angle=0.] {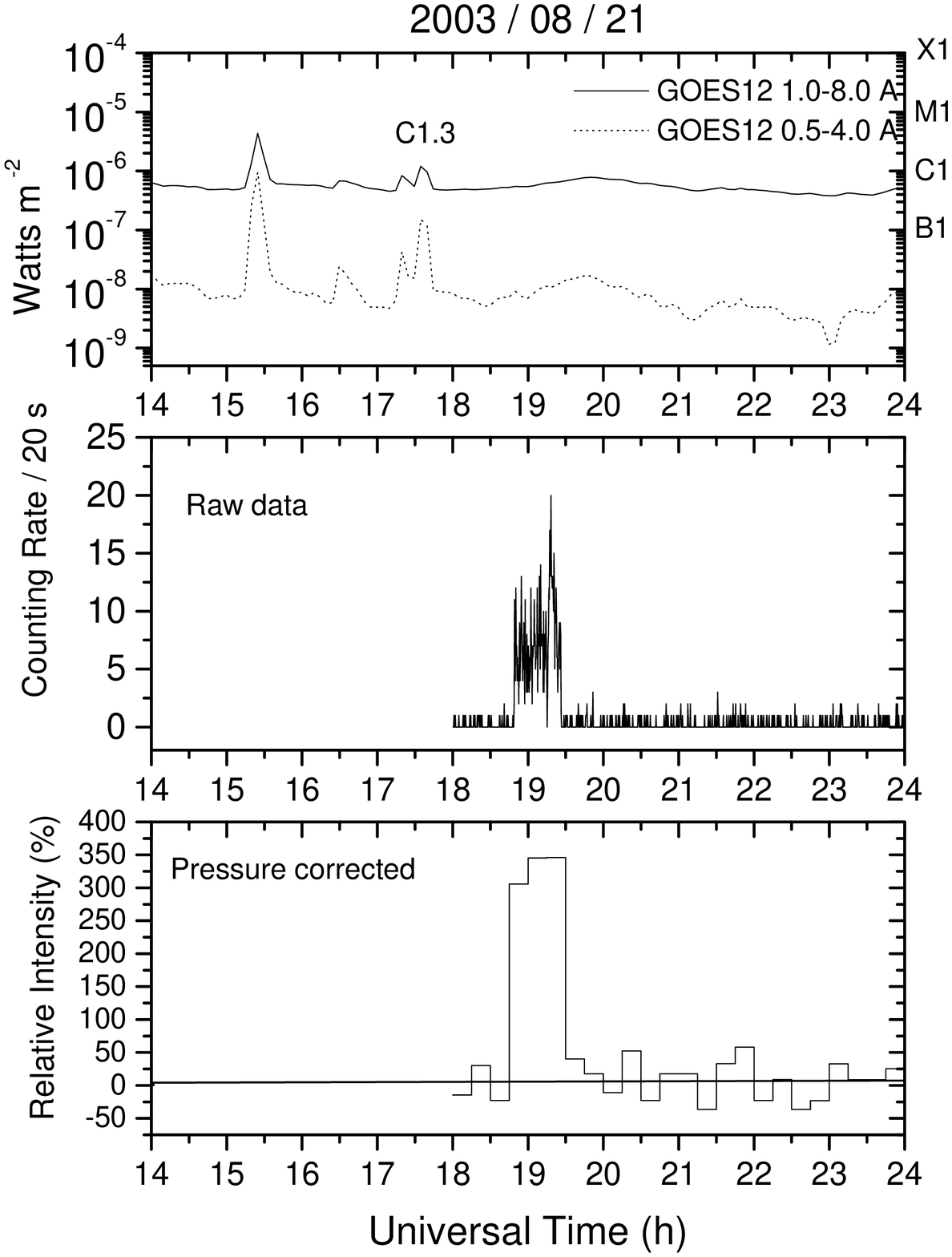}
\caption{Time profiles of the solar X-ray emission,
the TUPI raw data (counting rate), and the muon relative intensity (in the natural logarithmic representation) after pressure correction for the 2003/08/21  raster scan.}%
\end{figure}

\begin{figure}[th]
\includegraphics[clip,width=0.6
\textwidth,height=0.6\textheight,angle=0.] {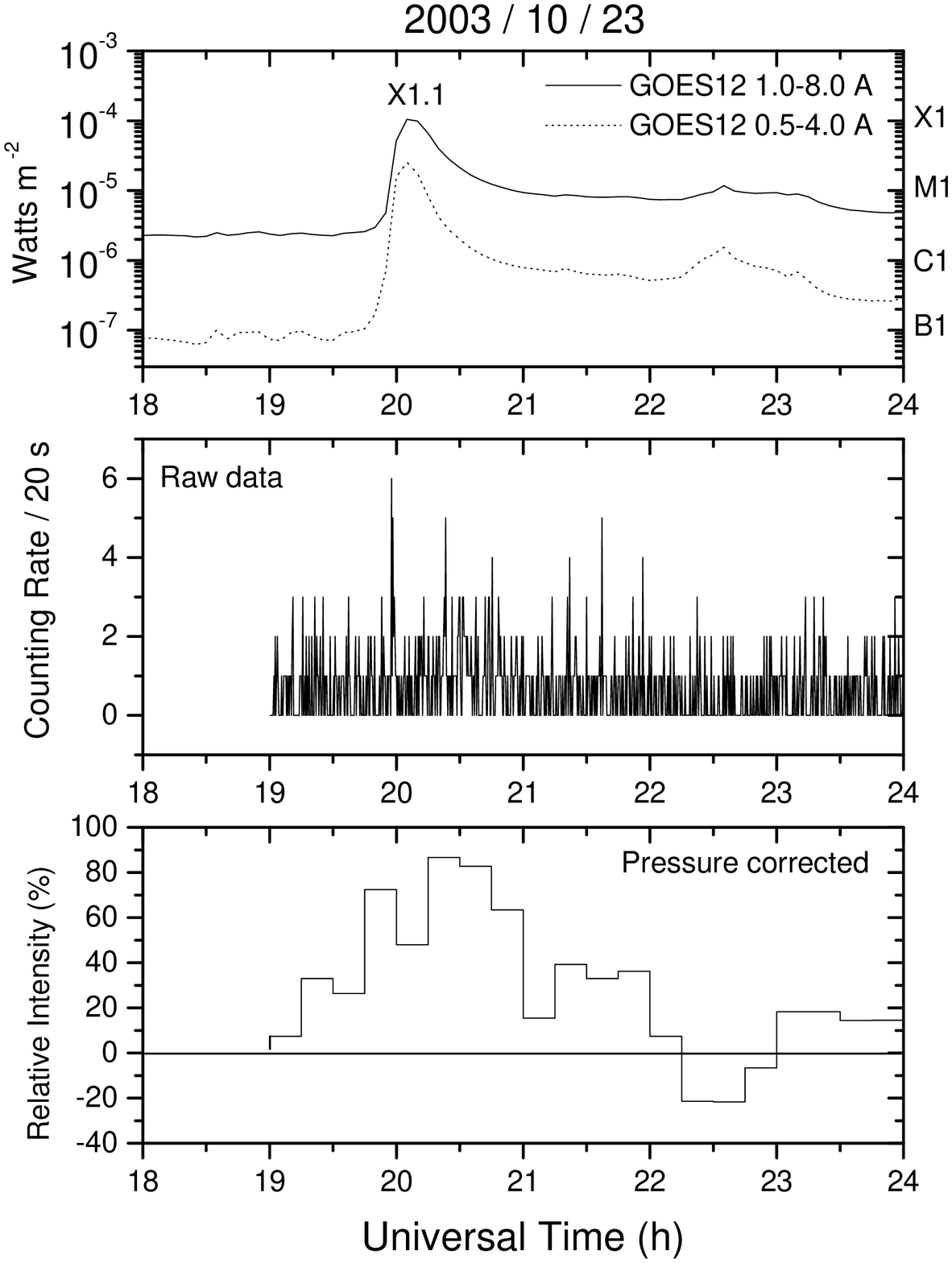}
\caption{Time profiles of the solar X-ray emission,
the TUPI raw data (counting rate), and the muon relative intensity (in the natural logarithmic representation) after pressure correction for the 2003/10/23 raster scan.}%
\end{figure}

\begin{figure}[th]
\includegraphics[clip,width=0.6
\textwidth,height=0.6\textheight,angle=0.] {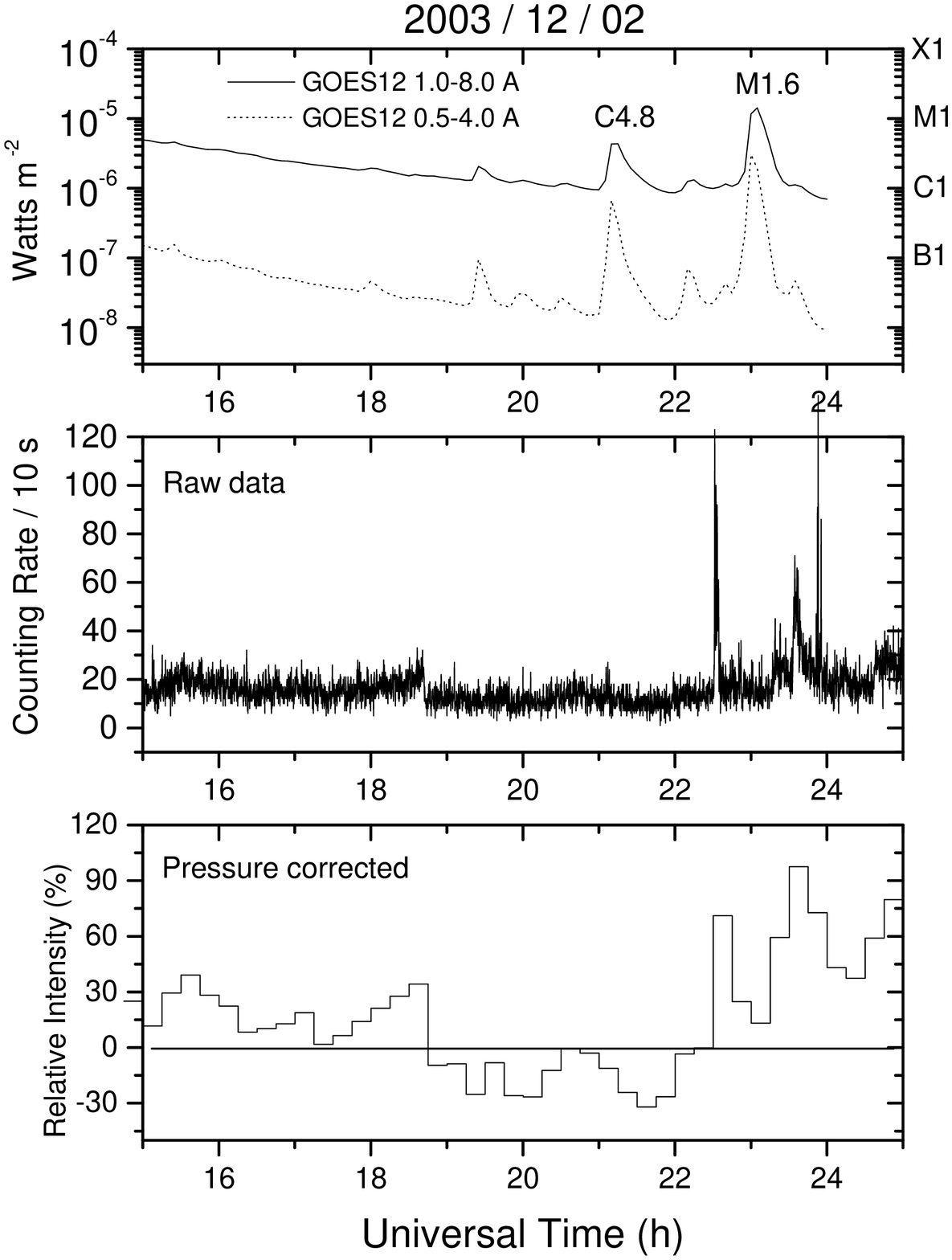}
\caption{Time profiles of the solar X-ray emission,
the TUPI raw data (counting rate), and the muon relative intensity (in the natural logarithmic representation) after pressure correction for the 2003/12/02  raster scan.}%
\end{figure}

\begin{figure}[th]
\includegraphics[clip,width=0.6
\textwidth,height=0.6\textheight,angle=0.] {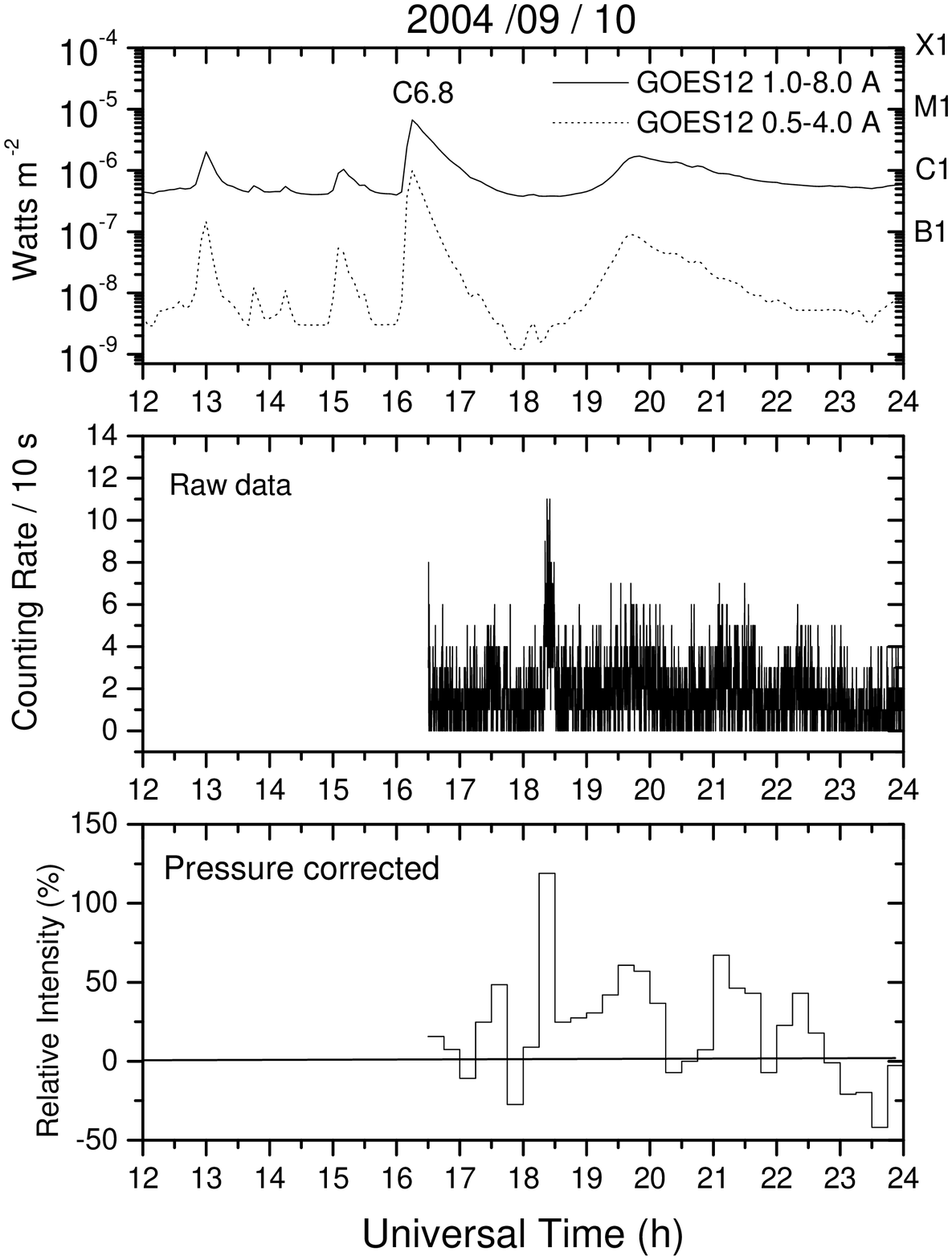}
\caption{Time profiles of the solar X-ray emission,
the TUPI raw data (counting rate), and the muon relative intensity (in the natural logarithmic representation) after pressure correction for the 2004/09/10  raster scan.}%
\end{figure}

\begin{figure}[th]
\includegraphics[clip,width=0.6
\textwidth,height=0.6\textheight,angle=0.] {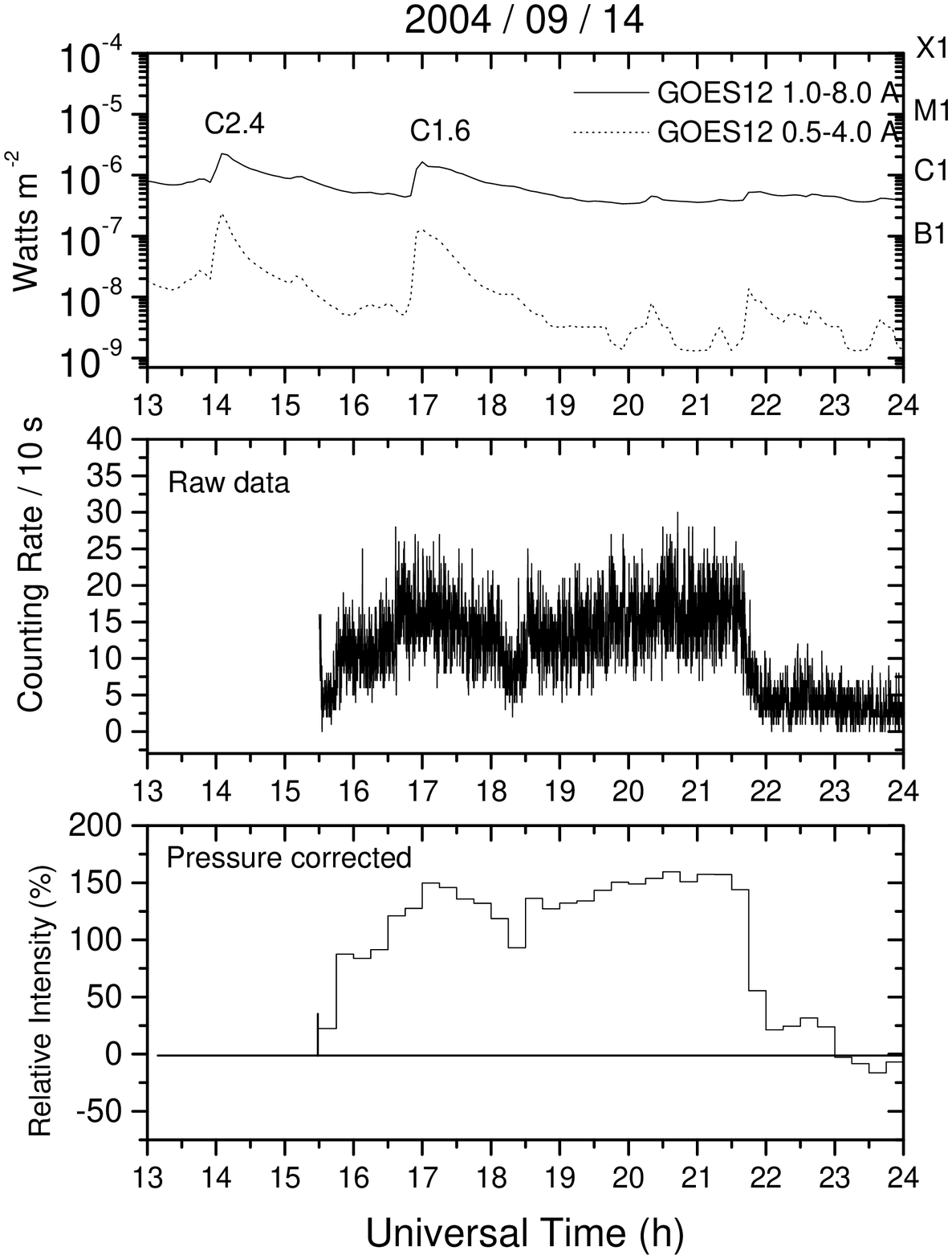}
\caption{Time profiles of the solar X-ray emission,
the TUPI raw data (counting rate), and the muon relative intensity (in the natural logarithmic representation) after pressure correction for the 2004/09/14  raster scan.}%
\end{figure}

\begin{figure}[th]
\includegraphics[clip,width=0.6
\textwidth,height=0.6\textheight,angle=0.] {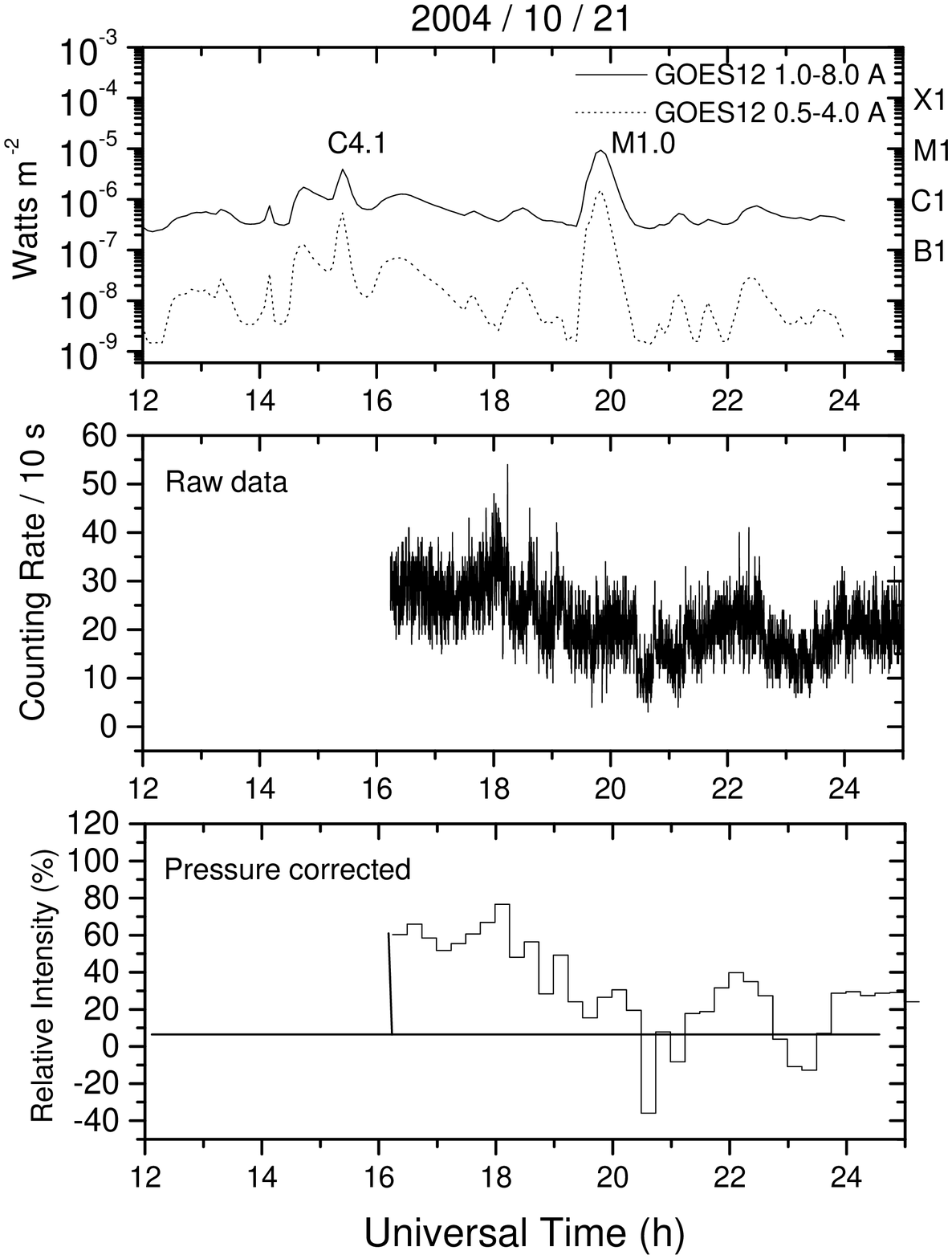}
\caption{Time profiles of the solar X-ray emission (background),
the TUPI raw data (counting rate), and the muon relative intensity (in the natural logarithmic representation) after pressure correction for the 2004/10/21  raster scan.}%
\end{figure}

\end{document}